\documentclass[12pt]{article}
\usepackage{epsfig}
\usepackage{amssymb,amsmath}
\newcommand{\beq}{\begin{equation}}
\newcommand{\eeq}{\end{equation}}
\newcommand{\bea}{\begin{eqnarray}}
\newcommand{\eea}{\end{eqnarray}}

\def\e2sig{e^{-2r\sigma}}

\begin{document}
\setlength{\baselineskip}{18pt}

\begin{titlepage}

\begin{flushright}
OCU-PHYS-460
\end{flushright}

\vspace*{10mm}

\begin{center}
{\bf\Large Anomalous Higgs Yukawa Couplings \\
and Recent LHC Data}
\end{center}

\vspace*{8mm}
\begin{center}
{\large 
Arindam Das$^{~a, b, c,}$\footnote{Electronic address: arindam@kias.re.kr}, 
Nobuhito Maru$^{~d,}$\footnote{Electronic address: nmaru@sci.osaka-cu.ac.jp} 
and Nobuchika Okada$^{~e,}$\footnote{Electronic address: okadan@ua.edu}
} 
\end{center}
\vspace*{0.2cm}
\begin{center}
${}^{a}${\it School of Physics, KIAS, Seoul 130-722, Korea}\\
${}^{b}${\it Department of Physics \& Astronomy, Seoul National University,\\ 1 Gwanak-ro, Gwanak-gu, Seoul 08826, Korea}\\
${}^{c}${\it Korea Neutrino Research Center, Bldg 23-312, Seoul National University, \\
Sillim-dong, Gwanak-gu, Seoul 08826, Korea}\\
${}^{d}${\it Department of Mathematics and Physics, Osaka City University, \\
Osaka 558-8585, Japan}\\
${}^{e}${\it 
Department of Physics and Astronomy, University of Alabama, \\
Tuscaloosa, Alabama 35487, USA} 
\end{center}
\vspace*{1cm}
\begin{abstract}
Very recently, the CMS collaboration has reported a search for the production 
  for a Standard Model (SM) Higgs boson in association with a top quark pair ($t \bar{t} H$) 
  at the LHC Run-2 and a best fit $t \bar{t} H$ yield of $1.5 \pm 0.5$ times the SM prediction 
  with an observed significance of $3.3 \sigma$.   
We study a possibility of whether or not this observed deviation can be explained by 
  anomalous Higgs Yukawa couplings with the top and the bottom quarks, 
  along with the  LHC Run-1 data for the Higgs boson properties. 
We find that anomalous top and bottom Yukawa couplings 
   with about $0-20$\% and $10-40$\% reductions from their SM values, respectively, 
   can simultaneously fit the recent CMS result and the LHC Run-1 data. 
\end{abstract}
\end{titlepage}

After the discovery of Higgs boson at the Large Hadron Collider (LHC) \cite{ATLAS, CMS}, 
   a next important task of the LHC experiments is to test whether the Higgs boson properties are consistent 
   with those predicted by the Standard Model (SM) or not. 
The LHC Run-1 data for Higgs boson properties from the combined analysis by the ATLAS and CMS collaborations 
   are summarized in Ref.~\cite{Higgs-Properties}. 
Although the current data show that the measured Higgs boson properties are more or less consistent 
  with the SM predictions, more precise measurements are necessary to conclude a consistency of the SM 
  or reveal a deviation from the SM predictions.

Very recently, the CMS collaboration has reported a search for the production 
  for an SM Higgs boson in association with a top quark pair ($t \bar{t} H$) 
  at the LHC Run-2 \cite{CMS-ttH}.  
Targeting the final states from the Higgs boson decays to $W W^*$ and $Z Z^*$, 
  the CMS collaboration has obtained a best fit $t \bar{t} H$ yield of $1.5 \pm 0.5$ times the SM prediction 
  with an observed significance of $3.3 \sigma$.   
This observed deviation from the SM prediction might be from anomalous Higgs couplings. 
If this is the case, we may consider two possibilities for the origin of the deviations. 
One is that the Higgs boson has anomalous top Yukawa coupling to enhance the Higgs boson 
   production cross section in association with a top quark pair.  
The other is that the branching ratio of the Higgs boson to $W W^*$ and $Z Z^*$ is enhanced from the SM case. 
Such an enhancement of the branching ratio is simply achieved if the bottom Yukawa coupling is anomalous 
   and smaller than the SM one. 
Since the decay width of the Higgs boson to a bottom quark pair dominates the total decay width,  
   the Higgs boson decay width to $W W^*$ and $Z Z^*$ is relatively enhanced 
   once the bottom Yukawa coupling is reduced.

Considering that the gauge structure of the SM has been very well established 
   from the electroweak precision measurements \cite{LEP1, LEP2}, 
   we focus on the Higgs boson Yukawa couplings to the SM fermions, 
   in particular, the Yukawa couplings for the third generation fermions.  
Since they are very small, the Yukawa couplings for the fermions in the first and the second generations 
   have negligible effects even if they are anomalous, except for flavor-changing anomalous Yukawa couplings. 
There are example models beyond the SM which predict anomalous Yukawa couplings of the physical Higgs boson, 
  such as the two Higgs doublet models \cite{Branco}, a 5-dimensional extension of the SM \cite{CHNOY} 
  in the Randall-Sundrum warped background \cite{RS}, and composite Higgs models \cite{comHiggs}.

In this paper, we study a possibility whether such anomalous Yukawa couplings can explain 
   the recent CMS observation of the enhanced $t \bar{t} H$ production cross section 
   along with the LHC Run-1 data for the Higgs boson properties.\footnote{
There are previous studies in the similar direction \cite{ttH_Pre} based on the LHC Run-1 data \cite{Higgs-Properties}, 
   where the top and bottom quark Yukawa couplings are measured to be $2\sigma$ away from the SM predictions. 
}    
We numerically analyze the effects of anomalous top and bottom (and tau) Yukawa couplings 
   on the $t \bar{t} H$ production, followed by the Higgs boson decays to $W W^*$ and $Z Z^*$, 
   as well as the signal strengths of Higgs boson decay modes for  
   $H \to \gamma\gamma, WW, ZZ, b\bar{b}, \tau \bar{\tau}$.  
Then, we identify a parameter region which is consistent with both the recent CMS result  
   and the LHC Run-1 data.

In the SM, a Yukawa coupling is given by
\bea
Y_f \bar{f} \Phi f = Y_f \bar{f} \langle \Phi \rangle f + \frac{Y_f}{\sqrt{2}} \bar{f} H f, 
\eea
   where $\Phi$ is the electric charge neutral component of the SM Higgs doublet field, 
   $H$ is the physical Higgs boson obtained by expanding $\Phi$ around its vacuum expectation value (VEV), 
   and $f$ represents an SM fermion.   
In the right-hand side, the first term is a fermion mass $m_f = Y_f \langle \Phi \rangle= Y_f v/\sqrt{2}$ 
  with $v\simeq 246$ GeV, and 
   the second term denotes the Yukawa coupling between the physical Higgs boson and the SM fermion. 
In the SM, the Yukawa coupling of the physical Higgs boson with the fermion is predicted 
    to be the ratio of the fermion mass to the Higgs VEV, $Y_f=m_f/v$. 
However, this fact is not necessarily true in some new physics models beyond the SM as mentioned above. 
In the following analysis, let us parametrize the deviations of top and bottom Yukawa couplings from the SM ones as
\beq
Y_t = c_t Y_t^{{\rm SM}}, \quad  Y_b = c_b Y_b^{{\rm SM}},
\eeq 
where $Y_t^{{\rm SM}}$ and $Y_b^{{\rm SM}}$ are the SM Yukawa couplings for the top and the bottom quarks, respectively, 
   and $c_t=c_b=1$ is the SM case.

Now we calculate the partial decay widths for various Higgs decay modes: 
\bea
\Gamma_{H \to gg} &=& c_t^2 \Gamma_{H \to gg}^{{\rm SM}}, \\
\Gamma_{H \to bb} &=& c_b^2 \Gamma_{H \to bb}^{{\rm SM}}, \\
\Gamma_{H \to \gamma\gamma} 
&=& \frac{\alpha_{em}^2 m_H^3}{256\pi^3 v^2} 
\left[ \frac{4}{3} c_t^2 F_{1/2}(m_H) + F_1(m_H) \right]^2, \\
\Gamma_{H \to \gamma Z} &=& 
\frac{\alpha_{em}m_W^2 m_H^3}{128\pi^4 v^4} \left( 1 - \left( \frac{m_Z}{m_H} \right)^2\right)^3 \nonumber\\ 
&\times &\left[ \frac{2}{\cos \theta_W} \left( 1- \frac{8}{3} \sin^2 \theta_W \right)c_t F_{1/2}(m_H) + F_1(m_H) \right]^2, \\
\Gamma_{H \to cc} &=& \Gamma_{H \to cc}^{{\rm SM}}, \\
\Gamma_{H \to \tau\tau} &=& \Gamma_{H \to \tau\tau}^{{\rm SM}}, \\
\Gamma_{H \to WW^*} &=& \Gamma_{H \to WW^*}^{{\rm SM}}, \\
\Gamma_{H \to ZZ^*} &=& \Gamma_{H \to ZZ^*}^{{\rm SM}}, 
\eea
where the partial decay widths in the SM are given by 
\bea
\Gamma_{H \to gg}^{{\rm SM}} &=& \frac{\alpha_s^2 m_H^3}{128\pi^3 v^2} (F_{1/2}(m_H))^2, \\
\Gamma_{H \to WW^*}^{{\rm SM}} &=& \frac{3m_W^4 m_H}{32\pi^3 v^4} G\left( \frac{m_W^2}{m_H^2} \right), \\
\Gamma_{H \to ZZ^*}^{{\rm SM}} &=& \frac{3m_Z^4 m_H}{32\pi^3 v^4} 
\left( \frac{7}{12} -\frac{10}{9} \sin^2 \theta_W + \frac{40}{9} \sin^4 \theta_W \right)
G\left( \frac{m_W^2}{m_H^2} \right), \\
\Gamma_{H \to bb}^{{\rm SM}} &=& \frac{3m_H m_b^2}{8\pi v^2} \left( 1- \frac{4m_b^2}{m_H^2} \right)^{3/2}, \\
\Gamma_{H \to cc}^{{\rm SM}} &=& \frac{3m_H m_c^2}{8\pi v^2} \left( 1- \frac{4m_c^2}{m_H^2} \right)^{3/2}, \\
\Gamma_{H \to \tau\tau}^{{\rm SM}} &=& \frac{m_H m_\tau^2}{8\pi v^2} \left( 1- \frac{4m_\tau^2}{m_H^2} \right)^{3/2}, 
\eea
and the loop functions are \cite{guide} 
\bea
F_{1/2}(m_H) &=& -2 \frac{4m_t^2}{m_H^2} \left[ 1- \left( 1 - \frac{4m_t^2}{m_H^2} 
{\rm arcsin}^2\left( \frac{m_H}{2m_t} \right) \right) \right], \\
F_1(m_H) &=& 2 + 3 \frac{4m_W^2}{m_H^2} + 3\frac{4m_W^2}{m_H^2} \left( 2 -\frac{4m_W^2}{m_H^2} 
{\rm arcsin}^2\left( \frac{m_H}{2m_t} \right) \right), \\
G(x) &=& 3 \frac{1-8x+20x^2}{\sqrt{4x-1}} {\rm arccos}\left( \frac{3x-1}{2x^{3/2}} \right) 
- \frac{1-x}{2x}(2-13x+47x^2) \nonumber \\
&& -\frac{3}{2} (1-6x+4x^2)\log x. 
\eea
In our analysis, we employ the center value of the Higgs boson mass $m_H=125.09$ GeV \cite{Higgs-Properties}
   and the center value of the combination of Tevatron and LHC measurements 
   of the top quark mass $m_t=173.34$ in GeV \cite{Mt}.

Signal strength ratios of the Higgs boson production via mainly gluon fusion and its decay are calculated from
\bea
\mu^{\gamma \gamma}
&=& \frac{\Gamma_{H\to gg} + 0.109~\Gamma^{SM}_{H \to gg}}{\Gamma^{SM}_{H \to gg} + 0.109~\Gamma^{SM}_{H \to gg}} 
\times \frac{BR(H \to \gamma\gamma)}{BR(H \to \gamma\gamma)_{SM}}, \\
\mu^{WW}
&=& \frac{\Gamma_{H\to gg} + 0.109~ \Gamma^{SM}_{H \to gg}}{\Gamma^{SM}_{H \to gg}+0.109~\Gamma^{SM}_{H \to gg}} 
\times \frac{BR(H \to WW^*)}{BR(H \to WW^*)_{SM}}, \\
\mu^{ZZ}&=& \frac{\Gamma_{H\to gg} + 0.109~ \Gamma^{SM}_{H \to gg}}{\Gamma^{SM}_{H \to gg} + 0.109~\Gamma^{SM}_{H \to gg}} 
\times \frac{BR(H \to ZZ^*)}{BR(H \to ZZ^*)_{SM}}, \\ 
\mu^{bb} &=& \frac{\Gamma_{H\to gg} + 0.109~ \Gamma^{SM}_{H \to gg}}{\Gamma^{SM}_{H \to gg} + 0.109~\Gamma^{SM}_{H \to gg}} 
\times \frac{BR(H \to bb)}{BR(H \to bb)_{SM}}, \\
\mu^{\tau\tau} &=& \frac{\Gamma_{H\to gg} + 0.109~ \Gamma^{SM}_{H \to gg}}{\Gamma^{SM}_{H \to gg} 
+ 0.109~\Gamma^{SM}_{H \to gg}} 
\times \frac{BR(H \to \tau\tau)}{BR(H \to \tau\tau)_{SM}}.  
\eea
Here, we have added $0.109~ \Gamma^{{\rm SM}}_{H \to gg}$,  
   which corresponds to the contributions of Higgs boson production 
   by the vector boson fusion channels (see Table~9 in \cite{Higgs-Properties}), 
  and other contributions to Higgs boson production have been neglected.

For the Higgs boson production associated with a top quark pair, followed by the Higgs boson decays to $WW^*$ and $ZZ^*$, 
   we evaluate its signal strength ratio to the SM prediction by 
\bea 
   \mu^{ttH}= c_t^2 \times \frac{BR(H \to WW^*)}{BR(H \to WW^*)_{SM}} 
                 = c_t^2 \times \frac{BR(H \to ZZ^*)}{BR(H \to ZZ^*)_{SM}}.  
\eea   
Since the partial decay widths to $WW^*$ and $ZZ^*$ are independent of $c_t$ and $c_b$, 
    $\mu^{ttH}$ with the Higgs boson decay to $WW^*$ is the same as that with the Higgs boson decay to $ZZ^*$.

\begin{figure}[t]
\begin{center}
\includegraphics[width=10.0cm,bb=0 0 298 294]{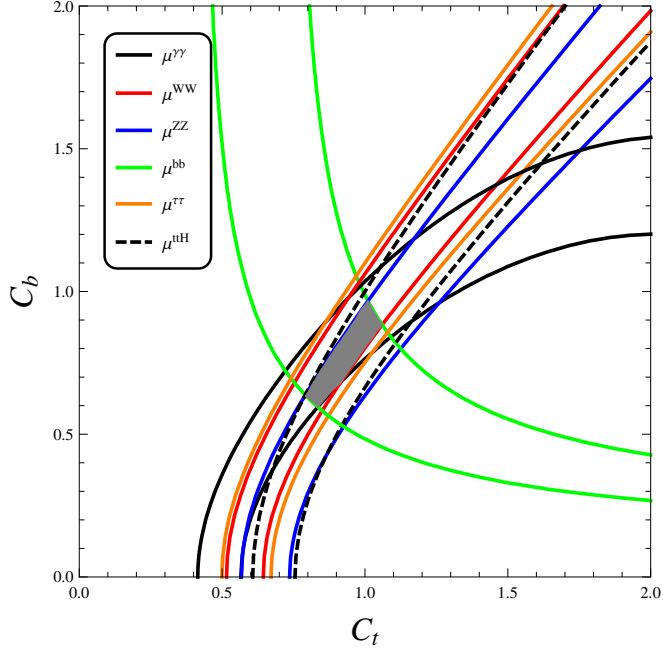}
\end{center}
\vspace*{-7mm}
\caption{
Signal strength ratios in $(c_t, c_b)$-plane. 
Here, we have used the constraints from the LHC Run 1 data for various modes (68\% confidence level): 
 $0.96  \leq  \mu^{\gamma \gamma} \leq 1.33$, 
 $0.93  \leq  \mu^{WW}  \leq1.27$, 
 $1.06  \leq  \mu^{ZZ}    \leq1.55$, 
 $0.43  \leq  \mu^{bb}     \leq 0.99$, and
 $0.89 \leq \mu^{\tau \tau} \leq1.35$  (see Table~13 in \cite{Higgs-Properties}).  
For the Higgs production associated with a top quark pair,  we have used the CMS result \cite{CMS-ttH}:  
 $1.0 \leq \mu^{ttH} \leq 2.0$. 
The (gray) shaded region simultaneously satisfies the LHC Run 1 data and the CMS result.     
}
\label{result1}
\end{figure}

Fig.~\ref{result1} shows our results in $(c_t, c_b)$-plane for various signal strength ratios 
   set by the LHC Run 1 data at 68\% confidence level \cite{Higgs-Properties} 
   and the CMS result for the production in association with a top quark pair, 
   followed by $H \to WW^*/ZZ^*$.   
The (gray) shaded region simultaneously satisfies the LHC Run 1 data and the CMS result, 
   where the anomalous Yukawa couplings range between $0.8 \lesssim c_t \lesssim 1 $ and $0.6 \lesssim c_b \lesssim 0.9$.  
It is an interesting observation that only the anomalous bottom Yukawa coupling of $c_b \simeq 0.9$  
    is sufficient to be consistent with all the LHC data.

\begin{figure}[t]
\begin{center}
\includegraphics[width=10.0cm,bb=0 0 298 294]{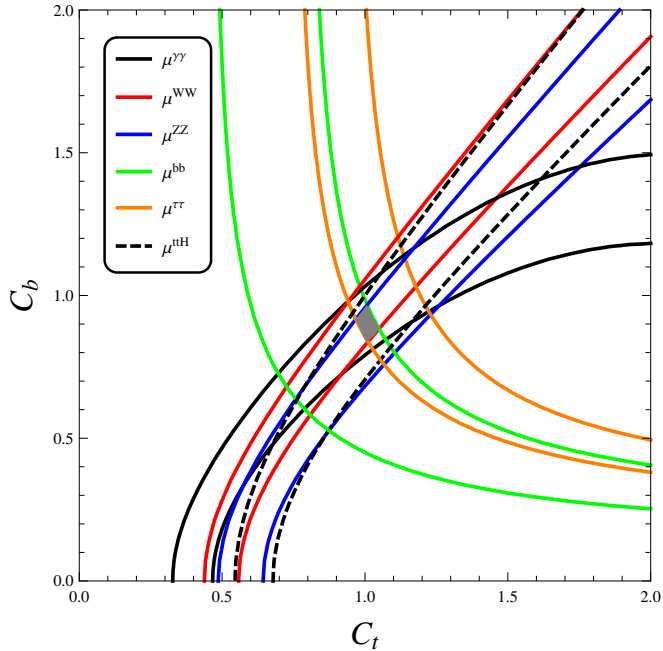}
\end{center}
\vspace*{-7mm}
\caption{
Same as Fig.~\ref{result1}, but we have here imposed the GUT relation of $c_b=c_\tau$. 
}
\label{GUT}
\end{figure}

One may be interested in considering a unification of bottom and tau Yukawa couplings, 
   which is usually predicted in Grand Unified Theories (GUTs). 
Although this relation holds only at a GUT scale, we apply it to our analysis, for simplicity. 
See, for example, Ref.~\cite{TeVGUT} on a GUT scenario with the coupling unification at the TeV scale.   
Let us introduce an anomalous coupling to tau Yukawa coupling, and parameterize it as 
\bea 
  Y_\tau=c_\tau Y^{\rm SM}_\tau,  
\eea
   and hence $\Gamma_{H \to \tau \tau} = c_\tau^2 \Gamma_{H \to \tau \tau}^{{\rm SM}}$.   
By imposing the GUT relation of $c_b=c_\tau$, we calculate signal strength ratios. 
Our results are shown in Fig.~\ref{GUT}. 
We find that $c_t \simeq 1$ and a suppressed bottom Yukawa coupling with $c_b\simeq 0.9$ 
   fit the LHC Run 1 data and the CMS result simultaneously.

Finally, we introduce a simple example model in which the physical Higgs boson couplings 
   to the SM fermions are indeed deviated from the SM ones. 
Let us consider the 5-dimensional SM in the Randall-Sundrum background (see, for instance, Ref.~\cite{CHNOY}). 
The original RS model \cite{RS} was proposed to provide a solution to the gauge hierarchy problem 
   by using five dimensional AdS space-time with a metric 
\bea
ds^2 &=& G_{MN} dx^M dx^N = e^{-2ky} \eta_{\mu\nu} dx^\mu dx^\nu - dy^2, \\
\eta_{\mu\nu} &=& {\rm diag}(+,-,-,-,-), 
\eea
compactified on an orbifold $S^1/Z_2$. 
Here, $x, y$ are the coordinates of four dimensional space-time and the fifth component, 
and $M,N (\mu, \nu)$ denote the indices for the five (four) dimensional space-time.

Suppose that the SM Higgs field $H(x,y)$ propagates in the bulk with the action  
\bea
S_H= \int d^4x dy \sqrt{-G} \left[ G^{MN} \partial_M H^\dag \partial_N H - \lambda (H^\dag H)^2 + m^2 H^\dag H \right], 
\eea
where the ordinary quartic Higgs potential is assumed. 
The equation of motion is obtained as
\bea
0 = -e^{2ky} \partial_\mu \partial^\mu H - \partial_y \partial^y H + 4k \partial^y H -(2\lambda H^\dag H - m^2) H.  
\eea
It is easy to see that the constant VEV is a solution of the equation of motion, $v=\sqrt{m^2/(2\lambda)}$. 
Expanding the Higgs field around the VEV in terms of Kaluza-Klein mode functions $f_n(y)$ as 
$H(x,y) = v + \tilde{H}(x,y)$ and $\tilde{H}(x,y) = \sum_n f_n(y) \tilde{H}_n(x)$
and plugging it back into the equation of motion, we find
\bea
\frac{d^2 f_n(y)}{dy^2} -4k \frac{d f_n(y)}{dy} - ( e^{2ky} m_n^2 - 2m^2) f_n = 0
\eea
where $- \partial_\mu \partial^\mu \tilde{H}(x) = m_n^2 \tilde{H}(x)$, and we have neglected the nonlinear term. 

The solution of this equation is given by Bessel functions, $J_\nu$ and $Y_\nu$, as 
\bea
f_n(y) = \frac{e^{2ky}}{\sqrt{N_n}} \left[ J_\nu(m_n e^{ky}/k) + b_\nu(m_n) Y_\nu(m_n e^{ky}/k) \right]
\eea
where $\nu=\sqrt{4+2m^2/k^2}$, $N_n$ is a normalization constant, and 
  $b_\nu(m_n)$ is a $y$ independent function determined by the boundary conditions. 
Now, our interest is on the lightest Kaluza-Klein mode ($n=1$). 
Noting that the Bessel functions are approximated for $m_1 e^{ky}/k \ll 1$ as follows, 
\bea
&&J_\nu (m_1 e^{ky}/k)  \simeq \left( \frac{m_1 e^{ky}}{2k} \right)^\nu \frac{1}{\Gamma(\nu+1)}, \\
&&Y_\nu (m_1 e^{ky}/k)  \simeq -J_{-\nu}(m_1 e^{ky}/k) \simeq  \left( \frac{m_1 e^{ky}}{2k} \right)^{-\nu} \frac{1}{\Gamma(-\nu+1)},
\eea
then, the mode function for the physical Higgs boson $f_1(y)$ has a nontrivial $y$-dependence,
while the Higgs VEV is a constant.  
Quarks and leptons are also bulk fields and can have nontrivial profiles in the fifth dimension. 
The four dimensional effective Yukawa couplings are given by the overlap integral 
of the quark and lepton mode functions and the mode function for the physical Higgs field. 
Therefore, Yukawa couplings between the physical Higgs boson and fermions are deviated from the SM ones.

In some extensions of the SM, such as two Higgs doublet models, Randall-Sundrum type models 
   and composite Higgs models, the relation between the Yukawa couplings of the physical Higgs boson 
   and the fermion masses in the SM can be violated. 
We have studied anomalous Higgs Yukawa couplings with the top and bottom quarks (and tau lepton) 
   in the light of the recent CMS result on the Higgs boson production associated with a top quark pair, 
   which shows a best fit $t \bar{t} H$ yield of $1.5 \pm 0.5$ times the SM prediction 
   with an observed significance of $3.3 \sigma$.   
We have found that the anomalous top and bottom Yukawa couplings can fit this CMS result   
   along with the LHC Run-1 data for the Higgs boson properties. 
In the most economic case,  only the anomalous bottom Yukawa coupling with around a 10\% reduction 
   is sufficient to be consistent with all the LHC data. 
The anomalous Yukawa couplings will be more precisely tested by the LHC Run-2 
   and the High-Luminosity LHC in the near future. 
The future International Linear $e^+ e^-$ Collider, once established, can provide us with a very precise 
   measurements of the Higgs boson properties, in particular, the bottom Yukawa coupling measurement 
   with an accuracy of ${\cal O}(1)$\%~\cite{ILC}.   
New physics beyond the SM may be revealed through a deviation of the Higgs boson couplings to SM fermions.

\subsection*{Acknowledgments}
We would like to thank Teruki Kamon for bringing the recent CMS result to our attention. 
We would also like to thank Yoshihiro Seiya and Kazuhiro Yamamoto for fruitful discussions on anomalous Higgs Yukawa couplings. 
The work A.D. is supported by the Korea Neutrino Research Center 
  which is established by the National Research Foundation of Korea (NRF) grant funded 
  by the Korea government (MSIP) (No.~2009-0083526). 
The work of N.O. is supported in part by the United States Department of Energy (No.~DE-SC0013680).


\end{document}